# Mobility as a Resource (MaaR) for resilient human-centric automation – a vision paper

**S. Travis Waller** (corresponding author)
Lighthouse Professor, Technische Universität Dresden, steven_travis.waller@tu-dresden.de

**Amalia Polydoropoulou**
Professor, University of the Aegean, polydor@aegean.gr

**Leandros Tassiulas**
Professor, Yale University, leandros.tassiulas@yale.edu

**Athanasios Ziliaskopoulos**
Professor, University of Thessaly, ziliasko@mie.uth.gr

**Sisi Jian**
Assistant Professor, Hong Kong University of Science and Technology, cesjian@ust.hk

**Susann Wagenknecht**
Jun-Professor, Technische Universität Dresden, susann.wagenknecht@tu-dresden.de

**Georg Hirte**
Professor, Technische Universität Dresden, georg.hirte@tu-dresden.de

**Satish Ukkusuri**
Reilly Professor of Civil Engineering, Purdue University, sukkusur@purdue.edu

**Gitakrishnan Ramadurai**
Professor, Indian Institute of Technology (IIT) Madras, gitakrishnan@iitm.ac.in

**Tomasz Bednarz**
Director, Strategic Researcher Engagement, NVIDIA, tbednarz@nvidia.com

## Abstract

With technological advances, mobility has been moving from a *product* (i.e., traditional modes and vehicles), to a *service* (i.e., Mobility as a Service, MaaS). However, as observed in other fields (e.g. cloud computing resource management) we argue that mobility will evolve from a *service* to a *resource* (i.e., "Mobility as a Resource", MaaR). Further, due to increasing scarcity of shared mobility spaces across traditional and emerging modes, the transition must be viewed within the critical need for ethical and equitable solutions for the traveling public (i.e., research is needed to avoid hyper-market driven outcomes for society). The evolution of mobility into a resource requires novel conceptual frameworks, technologies, processes and perspectives of analysis. A key component of the future MaaR system is the technological capacity to observe, allocate and manage (in real-time) the smallest envisionable units of mobility (i.e., atomic units of mobility capacity) while providing prioritized attention to human movement and ethical metrics related to access, consumption and impact. To facilitate research into the envisioned future system, this paper proposes initial frameworks which synthesize and advance methodologies relating to highly dynamic capacity reservation systems. Future research requires synthesis across transport network management, demand behavior, mixed-mode usage, and equitable mobility.

**Keywords:** Future Mobility, Transportation Systems, Modeling, Equity

## 1. Introduction

Human mobility needs as well as the foundation of human and freight movement, mobility infrastructure, are increasingly being disrupted by commoditization, digitization, automation (European Commission, 2022a), socially responsible goals (European Commission, 2022b; European Environment Agency, 2022), and novel business models (Polydoropoulou et al., 2020b) which are gradually leading to scarcity conditions (Cassetta et al., 2017; Dimitriou et al., 2020; Sprei, 2018). Further, within the scarce-capacity environment, there is a growing competition of shared space between humans utilizing increasingly diverse modes of transport (Abduljabbar et al., 2021; Miskolczi et al., 2021) (e.g., traditional modes, micromobilty, shared-mobility, ride-hailing, platform-based mobility, etc.) alongside automated technology (Burns, 2013; Narayanan et al., 2020). However, within this same dynamic of complex scarcity, novel forms of capacity are being generated through travelers carrying packages, citizen apps, on-demand transport, multimodal hubs or businesses bundling capacity as services, further transforming the operating conditions of the system (Rezende Amaral et al., 2018; Le et al 2019).

However, this transformation is not without its challenges. For instance, within the concept of the smart city, researchers have already been noting the utilitarian and ethical trade-offs related to the evolution of technology and human mobility with regard to the urban environment (Halpern and Günel, 2017; Kitchin, 2014; Sheller, 2018; Sheller and Urry, 2006). Further, as a focused technical framework, in environments of scarcity and technology-driven commoditization, there is a need to view limited shared entities as resources. Such as with water (Ibrahim, 2022) and computing (McKell et al., 1979) as a resource, we must begin to view mobility as a resource where we utilize technology to monitor, allocate and account for the smallest possible individual, atomic, units of the resource including dynamic property rights in time and space (i.e., specifying who is allowed to access, use and own the resource in a highly dynamic adaptive manner considering both human and automated movement).

Mobility transitioning into a resource is related to the broader social process of "resourcification" (Hultman et al, 2021; Corvellec et al, 2021) which examines the transformation of something into a resource where it had not traditionally existed as a resource. As noted by Hultman et al (2021), "… the process of turning things into resources is not always fair, sustainable, or well understood." Further, with scarcity, issues of equity, fairness, environmental justice, resilience and sustainability become even more pronounced and require focused attention on similarly high-fidelity, atomic, scales. Therefore, a new paradigm of Mobility as a Resource (MaaR) is presented for society's future mobility which radically expands on the long-term research in relation to automation, network control, adaptive equilibrium, demand forecasting, and dynamic capacity reservation.

As an analogy, the process of commoditization has demonstrated increased efficiency, adaptability, and robustness in other industries. As an example, computing has gone through this change process by transitioning from a product, to a service, to a resource (Zhan et al., 2015) (each step an increase in the commoditization of computing). As a resource, computing supply can be utilized at precisely the level needed without excess (in principle). Human mobility, which has additional complications due to the physical requirements of distant movement, is now making this transition from product to service (i.e., Mobility as a Service, MaaS) as demonstrated by the numerous MaaS companies and research contributions currently underway globally (Alonso-González et al., 2020; Hensher et al., 2021; Matyas and Kamargianni, 2019; Polydoropoulou et al., 2020a; Tirachini and Antoniou, 2020). As a note, while transitioning is stressed, just as MaaS has not entirely displaced the concept of car ownership, it is not asserted that MaaR would result in the complete elimination of previous paradigms (i.e., cloud computing has not rendered computer ownership as obsolete). However, it is envisioned that the transition to MaaR would make certain fundamental changes to the underlying systems that society relies upon for mobility.

Further, the concept of highly dynamic allocation of infrastructure for real-time traffic management has a lengthy history but is clearly growing rapidly with technological advancement. A few recent examples include research on dynamic bus lane allocation (Xie et al, 2012; Tsitsokas et al., 2021), intersection-based road and lane allocation (Alhajyaseen et al., 2017; Wu et al., 2021), auction-based managed lanes (Carlino et al., 2019; Levin et al., 2019; Rey et al., 2021) as well as control and optimization approaches for dynamic lane reversal (Levin et al., 2016; Levin et al., 2019). We assert that these pioneering examples will accelerate rapidly with a distinctive transition where all mobility supply will be viewed from a resource perspective at the highest fidelities of control and access as enabled by advances in automation, networked communication, data science, and applied AI.

## 2. The concept of Mobility as a Resource

An early precursor of the MaaR concept (Fajardo et al., 2011) examined the performance of a proposed intersection reservation system for automated vehicles as compared to traditional signalization. In this concept, small individual time-stamped road pavement sections can be reserved, utilized, and even traded between vehicles as a commodity (i.e., as a resource). In the subsequent years of urban transportation research, advancements continue being made to better exploit automation and network communication, which allow for the precise accountability and management of highly spatially and temporally disaggregate considerations necessary for high-fidelity resource management (Chen et al., 2021; Dresner and Stone, 2004; Fajardo et al., 2011; Guler et al., 2014; Peng et al., 2021). As a result, it is increasingly possible to envision a future where all shared mobility infrastructure can be commoditized and managed as a resource including interaction with non-vehicular modes such as cycling and walking by observing movements and ensuring automated components adapt quickly with appropriate space provisions.

MaaR views the evolution of transport in the form of ownership, management, and allocation of resources at the smallest possible scale to dynamically satisfy mobility demand while ensuring that the automated portions of the system are responsive to human movement. Essentially, the issue is proper prioritization of access (via dynamic reservation) to resources for individuals or entities who have needs for transportation services and/or partial or full sovereignty of resources that need to be engaged to accomplish the service demand. Reservation systems have received research attention (and in low fidelity solutions have existed throughout the history of transportation). Recent research has focused on highways (De Feijter et al., 2004; Ravi et al., 2007; Teodorovic et al., 2005), intersections (Carlino et al., 2013; Dresner and Stone, 2006, 2004; Fajardo et al., 2011; Vasirani and Ossowski, 2012), and some network (Chen et al., 2021) approaches.

MaaR envisions a much broader application of the reservation concept (as noted in Figure 1). It acknowledges that automated and non-automated travelers utilizing a broad means of modes will interact with a need for the reservations to be dynamically reallocated at microscopic scales, potentially well below what has been feasible historically. In addition, a key principle is that reservations may become invalidated (resulting in a series of cascading failures) when automated entities encounter unexpected human presence (i.e., with a cycle or e-scooter). Moreover, while travelers reserve physical (i.e., “hard”) capacity for their own movement (denoted by the lower section of Figure 1), they have the potential of providing “soft” capacity in the form of small package delivery. Such a system represents a significant novel user experience. Given how substantially user experience can change, there is a series of traffic flow modeling, network equilibration, and travel demand forecasting research required to understand the implications of such a radically differing travel experience. Finally, with substantially altered system-wide behavior, novel planning research is required to understand the impacts of emergent system dynamics on long-term strategic transport planning.

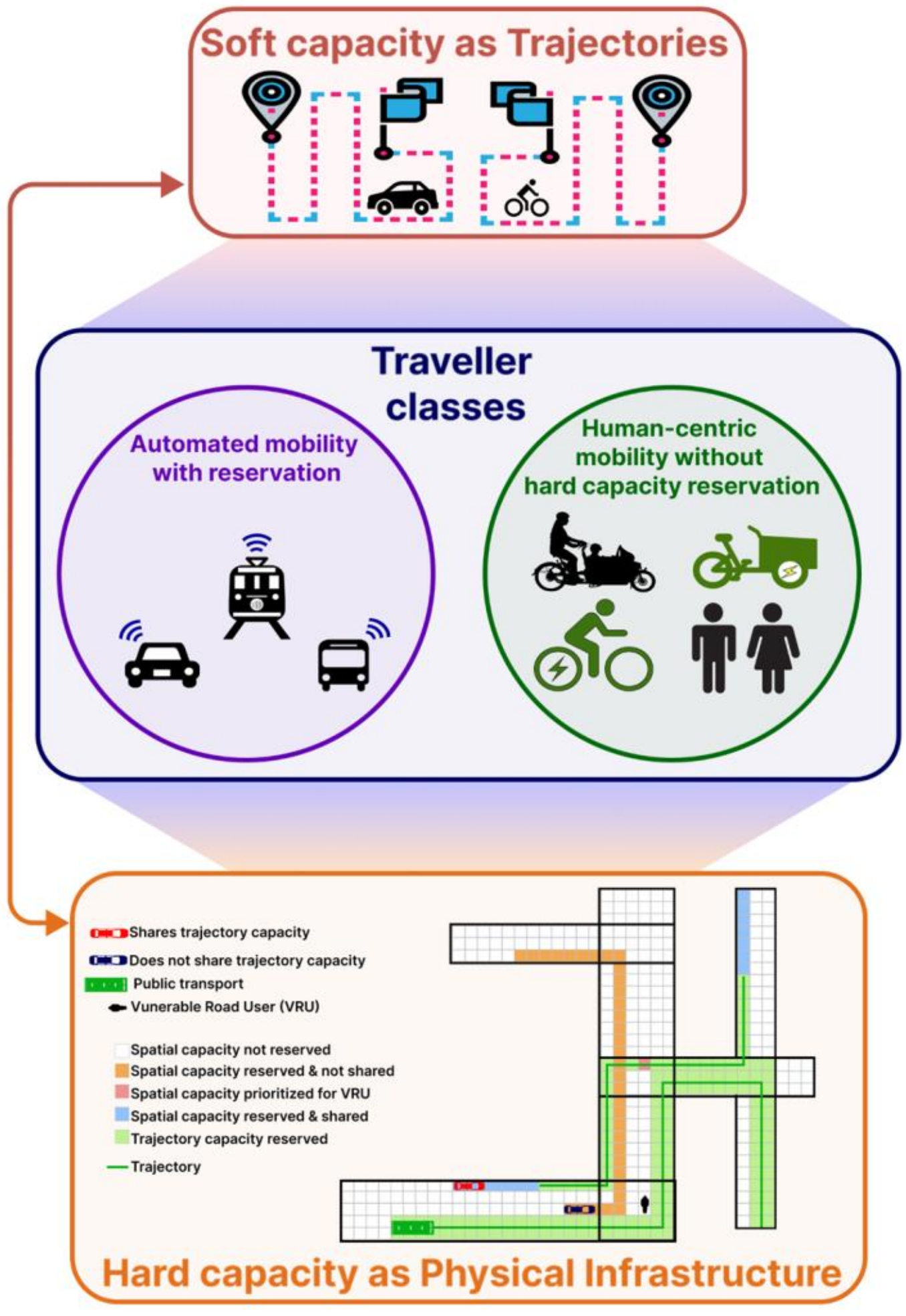

**Figure 1**: Research Concept of Dynamic Reservations of Resources for Multiple Classes of Travelers

## 3. Roadmap to Mobility as a Resource

To prepare for a future with MaaR, multiple novel but workable, concepts must be developed (e.g., identification of resource units across the mobility landscape, and understanding the preferences and roles of distinct stakeholders). While existing work facilitates articulating a vision, significant research is required to formulate, model, and explore the quantitative impact of design and policy decisions related to future MaaR deployments. One potential roadmap to MaaR includes the following 5 pillars:

### Pillar 1: Establishing a Future Framework for MaaR

The first step is to establish the framework(s) required for the function of Mobility as a Resource (MaaR). As noted, MaaR is a potential future mobility scenario where system users can reserve small segments of the physical infrastructure for a specified duration (and potentially generate novel 'soft' capacity in the process). The concept of reservation is clearly not new in transportation systems. Railways have been functioning based on this concept well before any modern systems model. Further, related concepts were used later in road infrastructure. To build a MaaR system, we need to expand the concept of small-unit dynamic reservation into a broader mobility context that embraces a representation of cascading failure and cross-system integration. Being a novel system, it poses numerous research questions. For instance, what mechanisms are suitable to control the reservations - A First Come First Serve (FCFS), an auction-based, or any other? On which platform should the system be based? Maybe a blockchain? Can the connected and non-connected systems coexist, and how do they interact- dedicated infrastructure for either system or can both share the same infrastructure? Even with dedicated infrastructure, how to control conflicting intersecting nodes? What should be the feasible mix of vehicles - 100% Autonomous Vehicles (AV) or a mix of AV and Human driven Vehicles (HV) is plausible? Which type of vehicles can share their capacity - only public transport and logistics providers or anyone in the system?

Critically, numerous potential implementation scenarios must be explored. Even with a focus on automation, human-centric mobility must be keenly represented keeping in mind that even for non-motorized travelers there is a high likelihood that they will interact with automated elements. The human-automation interface represents one of the essential components of the framework.

**Pillar 2: Automated/Mixed Traffic Flow and Control**

As noted, the MaaR concept views transportation resources divisible in space and time at scales that are fine enough to allow optimal performance benefits. In order to take advantage of this flexibility, dynamic control methodologies need to be developed that reshuffle mobility resources based on systems state measurements in real time. The recent development of the sensing and communication capabilities both at the vehicle and the infrastructure made such system state information available in real-time while controller decisions may be communicated to actuators in real-time as well. This evolution lead over the last decade the development of max-pressure stochastic control policies in transportation networks (Varaiya, 2013a, 2013b). The max-pressure approach was inspired by backpressure/max-weight policies in communication networks (Tassiulas and Ephremides, 1992) and have been considered initially for traffic control at intersections (Le et al., 2015; Levin, 2023; Li and Jabari, 2019; Manolis et al., 2018; Mercader et al., 2020; Robbennolt et al., 2022; Tsitsokas et al., 2021a, 2021b; Varaiya, 2013a, 2013b; Wongpiromsarn et al., 2012). Such approaches have the potential to replace the existing rigid schemes that follow the predefined allocation of transportation resources (Chow et al., 2020; Manolis et al., 2018). The finer real-time control in our MaaR framework will leverage max-pressure. However, the current max-pressure-based traffic control methods are designed to maximize the overall throughput. That is just scratching the surface of the wealth of optimization objectives that need to be quantified to address the vision of MaaR. Overall throughput maximization doesn't provide any guarantees for the performance of specific end-to-end traffic flows. In fact it is possible specific routes and corresponding flows to be throttled down in order to improve the overall throughput. We consider a multi-objective optimization framework where the performance of each individual traffic stream is represented as separate coordinate of the vector performance metric. Dynamic control methodologies will be pursued that can drive the performance vector to desirable feasible optimal points, leveraging techniques from Pareto optimization, game theory, and multi-agent systems as well as machine learning ((Feng and Tassiulas, 2022; Georgiadis et al., 2015; Koutsopoulos et al., 2014; Lei et al., 2020; Poularakis et al., 2021; Chen et al., 2022; Lei and Ukkusuri, 2023). Furthermore, fairness metrics will be developed that will ensure minimum performance guarantees to individual traffic flows that will alleviate uneven resource allocations and extreme inequalities among the mobility system participants including considerations of urgency (e.g., medical trips, work versus leisure, etc).

### Pillar 3: Adaptive Mixed Equilibrium for Network-based Metrics

MaaR foresees users competing/requesting for reservations of spatiotemporal resources of physical infrastructure for desired mobility. Reservations can be affected by disruptions and interactions with non-reserved forms of mobility, requiring real-time acquisition and re-allocations of resources. Such behavior will require modeling the individual and the resulting aggregate network behavior via equilibration to characterize and quantify the system properties, as this behavior underpins system-wide metrics. Typical examples include taxi market equilibrium (Qian and Ukkusuri, 2017) and electric mobility equilibrium (Chen et al., 2017).

In addition to hard capacity, another essential concept is the ability to share soft capacity, where users of hard capacity share their unused space with other passengers and/or freight. Parallels to the sharing economy as well as recent advancements in the research related to online resource allocation (Xi et al., 2023) can be leveraged. However, it is important to model the interactions between the reservation of hard and soft capacities within a single formulation. To develop such models, one can use the recourse trajectories obtained by solving the optimal tours for shared vehicles and solving stochastic shortest paths for the remaining vehicles in adaptive equilibrium or User Equilibrium with Recourse (UER) models (Unnikrishnan and Waller, 2009). However, significant further advancements are needed within the transport network modeling domain spanning the incorporation of equity considerations (e.g., (Najmi et al., 2023)), environmental justice (Duthie and Waller, 2008) as well as continued research into efficient techniques to enable the scale of implementation.

### Pillar 4: Traveler/Freight Behavior for MaaR

Future systems utilizing the MaaR vision will induce changes in travel behavior. However, as noted, technology is already rapidly disrupting traditional views of behavior. Even now, the increasingly complex mobility landscape challenges the standard behavioral models based on utility maximization (Ben-Akiva and Lerman, 1985; Manheim, 1979). More advanced modeling efforts have involved the development of hybrid choice models that include attitudes and perceptions of the decision-makers in the choice process, which have proven to represent the choice behavior of the individuals (Ben-Akiva et al., 1999) better. Already, to leverage the technologically-enabled rising potential of changing passenger and freight behavior, reward schemes as incentive for promoting multimodal choices have been developed (Figueiras et al., 2020; Polydoropoulou et al., 2018; Athena Tsirimpa et al., 2019; Athina Tsirimpa et al., 2019). Significant work has already begun to note the behavioral and framework implications of emerging mobility as a service insights (Hensher et al., 2023; Smith and Hensher, 2020). Moreover, to enhance the descriptive and predictive ability of the models, contemporary approaches are exploring the utilisation of machine learning techniques and neural networks (Hancock et al., 2020; Hillel et al., 2021; Salas et al., 2022; van Cranenburgh et al., 2022; Wong and Farooq, 2021, Wang et al., 2020) as alternatives or complements to Random Utility models.

Furthermore, considerable efforts to develop integrated spatial and multimodal transport planning tools to lead a sustainable transition to a new mobility era are currently under development (Polydoropoulou et al., 2022). However, all these models are basically supporting the status quo. Modeling the transition to MaaR requires the development of visionary new methods in combination with new technologies which will include the diverse portrayal of people, lifestyles, cultures, and mobility. MaaR is expected to affect various decisions, from strategic to tactical and operational decisions such as route choice or sharing modes, accounting for the fact that under MaaR, people, and a growing group of intermediaries that will rise with increased complexity costs, will be able to make selections dynamically.

By expanding the modeling capability related to traveler and freight for the noted future vision, key questions will need to be addressed as related to the relevant actors of the envisioned future landscape of MaaR (i.e., travelers, owners, providers, policy makers, etc.):

- Who are the decision-makers and what type of alternative decisions, not existing before, are to be made (i.e., willingness to use the reservation systems, buy or sell time/space or pay for different discrete elements in time and space, use the automated alternatives vs. conventional transport modes, use markets (auctions) or regulations to prioritize users and trip purposes, etc.)?
- How will we engage people to co-create and co-design the new transport solutions at such a high fidelity of space and time? How will alternatives be bundled? What incentives should be given to optimise user needs and transport system performance? How will we capture citizens' satisfaction with the MaaR solutions?
- What type of new technologies and information under MaaR will be available, and how will users adopt it over time? How will their behavior dynamically change?
- How will we address the diversity of the population (in terms of socioeconomic characteristics, vulnerabilities, attitudes and perceptions, lifestyles, cultures, build environment, availability of transport choices, physical accessibility, etc.) and their needs and requirements as well as how to ensure resilience and a just transition for all?
- How will we convince and incentivize people to change their behavior within MaaR to reduce the impacts of climate change and reduce the adverse effects of travel such as CO2 emissions? How do we measure social interaction effects?
- What type of prototype business and governance models should be developed to account for the new reality?

**Pillar 5: Synthesized Strategic Quantification for Societal Aims**

As MaaR is a future-oriented vision, it is unclear which trajectory may be taken by society. MaaR may be explored via a broad range of centralized or decentralized strategies (with the role of government a central consideration). Further, as additional shared space is incorporated into the system, there is the potential for a range of novel impacts on individual experience both on the positive and negative scale. This requires an integration to the growing body of knowledge on social welfare, equity, vulnerability and environmental justice as it pertains to the potential of human mobility broadly and MaaR impacts specifically. Ultimately, it is essential to evaluate whether MaaR improves social welfare. This implies evaluating instruments and mechanisms (e.g., markets, full planning, regulation, allocation of property rights: public or private, etc) for implementing MaaR with regard to welfare. Together with equity, vulnerability and environmental outcomes, this informs the future selection of appropriate instruments or mechanisms.

For instance, (Martens et al., 2019) have identified three fundamental components within a framework of distributive justice: the unit, scope, and shape of distribution. The unit of distribution pertains to the equitable allocation of benefits and burdens, such as public transport availability, accessibility to points of interest, or travel enjoyment. The notion of distributive equity in transportation has been noted as critical by academia, governments, and NGOs. However, the selection of equity metrics for transport planning represents a critical decision, and currently, there are standardization is highly limited for assessing equity performance (Adler et al., 2017; Behbahani et al., 2019; Caggiani et al., 2017; Guo et al., 2020, Mittal et al, 2023). Planners often choose equity measurements based on their specific objectives, resulting in significant divergence in terms of approach on measurement. Further, a particular project or policy may appear equitable when evaluated using one equity measure but may not be considered equitable when assessed using another (Behbahani et al., 2019; Ecola and Light, 2009).

The scope of distribution concerns the beneficiaries of these units, necessitating the disaggregation of the population into social groups to evaluate the distribution of benefits and burdens among them. Assessing equity entails categorizing the population into different groups based on various factors, such as income (El-Geneidy et al., 2016; Ricciardi et al., 2015), race (Karner and Niemeier, 2013), employment status (Lucas et al., 2003; Pyrialakou et al., 2016), gender (Dobbs, 2005; Rogalsky, 2010), immigrant status (Bennett and Shirgaokar, 2016), single parent status (Kramer and Goldstein, 2015; Pyrialakou et al., 2016), age (Jiao and Dillivan, 2013; Lucas et al., 2003), walkability (Rajat and Ukkusuri, 2023; Hamin and Ukkusuri, 2024) and disability (Al Mamun and Lownes, 2011; Hunter-Zaworski and Hron, 1999; Lucas et al., 2003).

The shape of distribution concerns the ideal conceptual allocation of benefits and burdens among the beneficiaries, often guided by moral principles (Jafino, 2021). Further, given the fundamental shifts in system performance with MaaR, there are significant behavioral implications as well. Therefore, behavioral welfare economics must be synthesized with the noted measures, indicators and considerations to aggregate the individual and societal well-being into a useful empirical welfare concept e.g., (Bernheim, 2008). Each of the aforementioned considerations requires novel mathematical modeling and social engagement with high degrees of consultation within democratic processes. As a result, the broad future-oriented vision of the implications of MaaR requires novel work spanning numerous disciplines particularly when it comes to the pricing, and prioritization of reservations.

Table 1 - A sample of applicable equity measures

| Equity Measurement | Formulation |
|---|---|
| Rawl's Egalitarian (RE) | $\text{RE} = \max \sum_{i=1}^{k} Y_i$ |
| Utilitarianism (U) | $\text{U} = \max \sum_{i=1}^{n} Y_i$ |
| Gini index (GINI) | $\text{GINI} = \frac{1}{2n^2\bar{Y}} \sum_{i=1}^{n} \sum_{j=1}^{n} \lvert Y_i - Y_j \rvert$ |
| Theil index (THEIL) | $\text{THEIL} = \frac{1}{n} \sum_{i=1}^{n} \left( \frac{Y_i}{\bar{Y}} \log\log \frac{Y_i}{\bar{Y}} \right)$ |
| Atkinson index (ATK) | $\text{ATK} = \begin{cases} 1 - \left[ \frac{1}{n} \sum_{i=1}^{n} \left( \frac{Y_i}{\bar{Y}} \right)^{1-\epsilon} \right]^{\frac{1}{1-\epsilon}}, \epsilon \neq 1 \\ 1 - \frac{1}{\bar{Y}} \left( \prod_{i=1}^{n} Y_i \right)^{\frac{1}{n}}, \epsilon = 1 \end{cases}$ |
| Sadr's theory of Justice (SADR) | $\text{SADR} = \begin{cases} \max \sum_{i=1}^{n} Y_i; \\ s.t\ Y_i > m1 \times Y_j, \forall i, j \\ \sum_{i,j} \frac{Y_i - Y_j}{2n^2\bar{Y}} < m2 \end{cases}$ |
| Relative mean deviation (RMED) | $\text{RMED} = \frac{1}{n} \sum_{i=1}^{n} \lvert \frac{Y_i}{\bar{Y}} - 1 \rvert$ |
| Mean log deviation (LDEV) | $\text{LDEV} = \frac{1}{n} \sum_{i=1}^{n} \lvert \log\log Y_i - \log\log \bar{Y} \rvert$ |

Where Y is a measure of transport-related accessibility or welfare; k is the number of protected groups; n is the number of observed groups; m1, m2 are gap parameters; ε in ATK are positive parameters of addressing inequity aversion.

## Concluding Remarks

Mobility as a Resource (MaaR) represents a vision for future human and freight mobility that could be an inevitable consequence of the increasing power of commoditizing technology to account for ever-smaller units of shared space and time in terms of ownership, access and usage. This future vision has the potential of unlocking exceptional system efficiency and enabling a broader view of higher-order applications built upon mobility resources but it also has the potential of incurring substantial impacts on individuals requiring close scrutiny of equity and environmental justice concerns.

The transition to this future is underway. As discussed in this paper, the vision proposes that mobility will continue its transition from a *product* (i.e., traditional modes and vehicles), to a *service* (i.e., Mobility as a Service, MaaS which is the current state of transport research and emerging practice) to a *resource* (MaaR, the eventual future state). This transition has a broad range of potential trajectories from highly centralized to highly decentralized with extremely diverse potential impacts on individual travelers. Multiple domains of ongoing research must be integrated to consider this evolving system so that it can be appropriately shaped towards a transition that is efficient, resilient and, most critically, societally-oriented in an equitable manner.

Moreover, the potential of MaaR to assist with the critical social goals of sustainability, equity and resilience relies heavily on the implementation trajectory that is adopted. Within various implementation trajectories there exists both ***potential*** but also ***emerging challenges***. For instance, in principle, given the highly dynamic nature of MaaR, resilience could be enhanced by a more rapid recovery following system disruptions. However, MaaR may also result in an exceptionally more complex system where cascading failures reveal much more intricate patterns of dependency. Further, MaaR could enable more personalized mobility where socio-demographic considerations might be better accounted for resulting in a more equitable distribution of access. However, if MaaR relies heavily on monetary markets for reservation assignment, it could worsen mobility equity. Similarly, for sustainability, MaaR could be beneficial if appropriate metrics for environmental impact and energy usage are incorporated fully thereby prioritizing the most sustainable modes and activities (i.e., greater overall movement of people with sustainable levels of impact and resource usage). However, if sustainability is not explicitly incorporated into the reservation allocation methodology, MaaR could even worsen sustainability by enabling even greater movements that result in unsustainable levels of impact. As a result, research is needed within each of the noted considerations to explore the eventual implementation trajectories that best align to societal goals.

In particular, **data science** research is vital to ensure positive outcomes. Our already substantial sets of mobility data will continue to grow alongside the same technology which is enabling MaaR thereby continuing to increase the applicability of data science techniques. In addition, the required research must be highly cross-disciplinary by incorporating the vast range of research in travel behavior, traffic flow, and network science (to name only a few) but fully integrated with quantitative data-driven insights.

Ultimately, from this research, new **policy tools** are required that exceed the capabilities of current state-of-the-art. In addition to policy tools, as with the entire field of automated mobility, significant work is required spanning governance, law and regulation which will vary by region. While difficult, the mobility field has found ways throughout its history of devising feasible solutions that eventually overcome regulatory complexities. Further, with new policy tools and regulatory frameworks, the broad range of scenarios previously identified can begin to come into view which could enable an implementation trajectory for the MaaR vision which aligns best for the broader society.

However, there are clear limitations to the MaaR vision. Most critically, it is highly reliant on the continued evolution of technology to support the fine-grained discretization of shared capacity at very small units of space and time. Further, as noted, the transition period is highly uncertain with numerous potential scenarios already envisioned but countless more to be explored. While the transition period may be substantial, given the potentially broad impact of MaaR, numerous lines of research should be underway urgently to help shape its impact into the future thereby ensuring positive outcomes for societal mobility.

## Declarations

### Ethical Approval
Not applicable

### Competing interests
None

### Authors' contributions
All authors reviewed the entire manuscript with comments, additions and revisions. Specific contributions include: S.T.W led overall vision formation and in particular contributed to pillar 3 with regard to adaptive network modeling. A.P. led vision on pillar 4 with regard to behavioral modeling implications. L.T. led vision on pillar 2 with regard to system control implications. A.Z. specifically contributed to pillars 3 and 4. S.J. specifically contributed to pillar 4 with additional contributions spanning the current state of Mobility as a Service (MaaS). S.W. contributed with regard to social engagement and broader societal implications. G.H. contributed in terms of the economic implications of the MaaR vision. S.U. contributed to Pillars 2 and 3. G.R. contributed to the framing of the resource concept versus services and broader choice modeling. T.B. contributed in terms of broader technological implications and capability.

### Funding
Not applicable

### Availability of data and materials
Not applicable